\documentstyle[aps,prb,twocolumn]{revtex}
\input{epsf}
\newlength{\colwi}
\begin{document}
\twocolumn[\hsize\textwidth\columnwidth\hsize\csname@twocolumnfalse\endcsname
\draft
\preprint{\today}
\title{
Quantum-Classical Correspondence in Energy Space: Two
Interacting Spin-Particles
}

\author{F.Borgonovi$^{[a,b,c]}$, I.Guarneri$^{[d,b,c]}$,
F.M.Izrailev$^{[d,e,f]}$ }
\address{
$^{[a]}$Dipartimento di Matematica, Universit\`a Cattolica,
via Trieste 17, 25121 Brescia, Italy  \\
$^{[b]}$Istituto Nazionale di Fisica Nucleare, Sezione di Pavia,
via Bassi 6, 27100 Pavia, Italy\\
$^{[c]}$Istituto Nazionale di Fisica della Materia, Unit\`a di Milano,
via Celoria 16, 22100 Milano, Italy\\
$^{[d]}$  Universit\'a di Milano, sede di Como,
via Lucini 3, 22100 Como, Italy\\
$^{[e]}$Budker Institute of Nuclear
Physics, 630090, Novosibirsk, Russia\\
$^{[f]}$Instituto de Fisica, Universidad Autonoma de Puebla, Apdo.
Postal J-48, Puebla 72570, Mexico\\
}

\maketitle
\begin{abstract}

The Hamiltonian conservative system of two interacting particles has
been considered both in classical and quantum description. The quantum 
model has been realized using a symmetrized two-particle basis reordered 
in the unperturbed energy. Main attention is paid to the structure of 
chaotic eigenfunctions (EF) and to the local spectral density of states
(LDOS). A remarkable correspondence has been found for the shapes of EF 
and LDOS in the energy representation, to their classical counterparts.
Comparison with the Band Random Matrix theory predictions has revealed 
 quite significant differences which are due to dynamical nature of the 
model. On the other hand, a partial agreement is found by inserting randomness
{\it `` ad hoc ''} in the dynamical model for two-body matrix elements.
This shows that, at least for small number of particles, care must be taken 
when classical correlations are neglected. The question of quantum 
localization in the energy space is discussed both for dynamical and 
random model.

\end{abstract}
\pacs{PACS numbers:
05.45+b}]

\section{Introduction}

Quantization of classically chaotic systems has been addressed, from the
very beginning, both to conservative and time--dependent systems. In the
latter case the important phenomenon of dynamical localization was
discovered, connecting a classical quantity, the diffusion rate, to the
quantum localization length of the correspondent equilibrium distribution 
\cite{CIS81}. Instead, in the case of conservative systems, important steps
have been done in establishing some distinctive features which mark a
quantum chaotic system from a integrable one: let us mention for instance,
the non-Wigner-Dyson statistics of neighboring level spacings, or the
scarring of eigenfunctions along some classical periodic orbits. However,
the possibility of quantum localization effects in such systems has been
scarcely explored until recently, when a clue in this direction was found 
\cite{CCGI96} by investigating a particular class of random models: the
Wigner Banded Random Matrices (WBRM) ensemble.

For such an ensemble, whose introduction dates back to Wigner himself \cite
{W55}, it is possible to obtain a series of results which allow for a
definition of quantum localization within the classical energy surface.
These results, when extended to Hamiltonian systems, would impose severe
quantum limitations on the behavior of classical ergodic systems. The
important result is that room is left for quantum localization and this can
obtained directly from the knowledge of local spectral density of states
(LDOS) and eigenfunctions (EF). Quite surprising, both quantities have well
defined classical limits (see Ref. \cite{CCGI96}), which,
generally speaking, have received scarce attention before now.

On the other hand, in order to acquire physical relevance, it is clear that
such results should be extended to real physical systems, where the origin
of randomness is purely dynamical. This we do, in this paper, by considering
a classically chaotic two--interacting spin system with a finite Hilbert
space. Our purpose is analyzing the structure of eigenstates and of LDOS,
and comparing it with expectations based on previous Random Matrix studies
and with their classical counterparts.

First of all we find that, when written in the eigenbasis of two
non-interacting particles, reordered in the unperturbed energy, the
Hamiltonian matrix has an overall banded structure. About the shape of
eigenfunctions and the LDOS, we find that our quantum results, in the
average, follow the behavior of similar quantities computed from WBRM 
 only approximately at best. Nevertheless, they follow
remarkably well the behavior of their classical analogs, that we actually
compute in the present paper.

On the other hand, the correspondence with Random Matrix Theories (RMT) is
restored on artificially randomizing our Hamiltonian. The lesson we draw
from this result is that, although RMT quite well reproduces fluctuation
properties of spectra of real chaotic Hamiltonians, some correlations are
missing in their structure, which are essential in giving the correct
semiclassical behavior when detailed questions about the structure of
eigenfunctions are asked. It is of course possible that a better
correspondence with RMT will be restored with systems with a larger number
of particles; for the time being, however, our results appear to indicate
that caution is needed in carrying over results from RMT to Hamiltonians
which have a smooth, well defined classical limit.

\section{The model}

\label{sec:level1}

The model has been proposed and widely investigated in \cite{FePe}. Here we
review few fundamental facts about its classical and quantum behavior. It
describes two coupled rotators, with angular momentum $\vec L$ and $\vec M$
with the following Hamiltonian: 
\begin{equation}
\label{ham1}H=A(L_z+M_z)+BL_xM_x
\end{equation}

It may be used to describe the interaction of quasi-spins in nuclear physics
or pseudo--spins in solid state systems. Choosing $A^{-1}$ as the unit of
time and $A B^{-1}$ as the unit of angular momentum, it can be written as $H
= H_0 + V$ where $H_0 = L_z + M_z$ and $V = L_x M_x$. The constants of
motion are $H=E$, $L^2$ and $M^2$.

Fixing the values of $L^2$ and $M^2$ it can also be shown\cite{FePe} that
the total energy must be bounded: 
\begin{equation}
\label{boun}E^2 \leq E^2_{max} = (L^2+1) (M^2+1)  
\end{equation}
for $LM >1$.

It is worth to mention that in this form the dynamical variables $\vec{L}, 
\vec{M}$ are not canonical. On the other side, the usual Hamiltonian form,
with the canonical variables $q_i, p_i$, $i=1,2$ can be recovered by means
of the following transformation :

\begin{equation}
\label{trasfo}
\begin{array}{ll}
L_x = & \sqrt{L^2 - p_1^2} \cos q_1 \\ L_y = & \sqrt{L^2 - p_1^2} \sin q_1
\\ L_z = & p_1 \\ 
M_x = & \sqrt{M^2 - p_2^2} \cos q_2 \\ M_y = & \sqrt{M^2 - p_2^2} \sin q_2
\\ M_z = & p_2 
\end{array}
\end{equation}

keeping $L^2$ and $M^2$ as constants\cite{FePeQ}. In these variables the
Hamiltonian reads: 
\begin{equation}
\label{hamnew}H = p_1 + p_2 +\sqrt{ L^2 - p_1^2} \sqrt{ M^2 - p_2^2}\cos q_1
\cos q_2  
\end{equation}

The analysis of the surfaces of section reveals a large number of regular
trajectories covering invariant tori when $L^2,M^2$ are both very small or
very large\cite{FePe}. To simplify the problem we set $L=M$. In such a case
the most interesting situation occurs when $1<L<10$ where, depending on the
energy value $E$, regular and chaotic regions coexist. Typically when $|E|$
is close to $E_{max}=L^2+1$ trajectories are regular while for $E\simeq 0$
islands of stability become very small and chaotic motion dominates.

Quantization follows standard rules, and angular momenta are quantized
according to the relations $L^2 = M^2 = \hbar^2 l(l+1)$ where $l$ is an
integer number. Therefore, for given $l$ the Hamiltonian is a finite matrix,
and the semiclassical limit is recovered in the limit $l \to \infty$ and $%
\hbar \to 0$ keeping $L^2$ constant.

In our approach the Hamiltonian is represented in the two--particles basis $%
\vert l_z, m_z \rangle$ where the matrix elements have the form,

\begin{equation}
\label{himp}\langle l_z^{\prime}, m_z^{\prime}\vert H_0 \vert l_z, m_z
\rangle = \delta_{m_z, m_z^{\prime}} \delta_{l_z, l_z^{\prime}} \hbar (l_z
+m_z)  
\end{equation}

and

\begin{eqnarray}
\langle l_z', m_z' \vert V   \vert  l_z,  m_z \rangle
= { {\hbar^2}\over {4}}  \delta_{m_z, m_z' \pm 1}   \delta_{l_z, l_z' \pm 1}
&\times 
\nonumber  \\
\sqrt{ (l+l_z) (l-l_z+1) (m+m_z) (m-m_z +1) } & 
\label{hint}
\end{eqnarray}
with $l_z, m_z$ integers, $-l \le l_z,m_z \le l$.

\begin{figure}
\centerline{\epsfxsize=0.7\colwi\epsffile{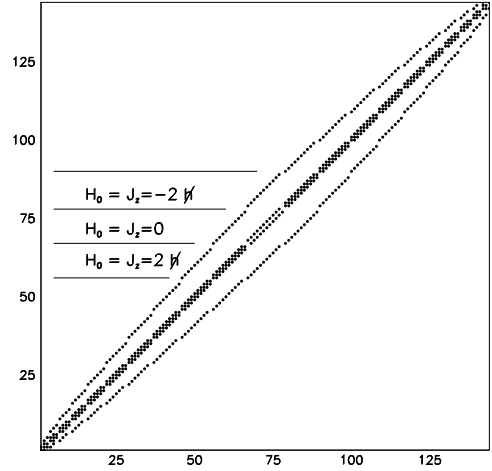}}
\caption{Structure of the Hamiltonian matrix in the symmetrized basis
for $l=m=11$, $(N=144)$ and $J_z$ even. Dots indicate
off-diagonal elements different
from zero. The bandwidth $b$ is maximal at the center where  $b=2l+1$.
Few different $H_0$ shells are shown by the horizontal
 lines.}
\label{figband}
\end{figure}

The $z$--component of the total angular momentum $J_z = L_z + M_z$ (which is
the same as 
the unperturbed Hamiltonian $H_0$) obeys the selection rules $\Delta J_z =
0, \pm 2\hbar$, so the subspace spanned by the states with odd $J_z/\hbar$
can be separated from that with $J_z/\hbar$ even (there are no matrix
elements for the transition between them). In what follows, we fix $%
J_z/\hbar = H_0/\hbar $ even. As a result, the matrix describing the
Hamiltonian has a dimension $N = 2 l^2 + 2 l + 1 $. We have also to take
into account the symmetry degeneracy with respect to the exchange of
particles. Below, we consider only symmetric states.

Let us now explain how the Hamiltonian matrix is constructed. Once $l$ is
fixed,  there are $2l+1$ single-particle levels $\langle -l \vert, \langle
-l +1\vert,\ldots, \langle -1 \vert, \langle 0 \vert, \langle 1
\vert,\ldots, \langle l-1 \vert, \langle l \vert$. The ground state is
represented by two particles in the lowest single-particle level, which we
label as $\langle -l, -l \vert$; it has an unperturbed energy $E_0 = -2 l
\hbar$. The first excited state is doubly degenerate and the two eigenstates
having the same energy $E_1 = (-2l+2) \hbar $ are $(\langle -l,-l+2 \vert +
\langle -l+2, -l \vert)/\sqrt{2}$ and $\langle -l+1, -l+1 \vert$. The
former state 
corresponds to one particle in the single-particle level $%
\langle -l+2\vert$ and the other in the single-particle ground state $%
\langle -l \vert$.
The latter state, to two particles in the single-particle level $%
\langle -l+1 \vert$. 
We call $H_0$-shell
the set of states having the same value of unperturbed
energy.

It is easy to prove that in the symmetrized basis each shell with 
$H_0$ fixed and even 
has a degeneracy $p= l+1 - \vert H_0 \vert / 2 \hbar$, and the
dimension of the Hamiltonian matrix is $N =(l+1)^2$ due to the relation 
$$
\sum_{H_0/2 = -\hbar l}^{\hbar l} ( l+1 - {\frac{{\vert H_0\vert} }{{2\hbar}}%
} ) = (l+1)^2.%
$$
We then reorder the Hamiltonian matrix according to the increasing
unperturbed energies and we call $\vert {\bf n} \rangle$ the resulting
two--particles symmetrized ordered basis.  As a result, the off-diagonal
matrix elements are symmetric with respect to the two main diagonals $%
H_{n,n} = \langle {\bf n} \vert H \vert {\bf n}\rangle $ and $H_{N-n,N-n}$.

Diagonal matrix elements are constructed from the unperturbed Hamiltonian $%
H_0$; they are given by the eigenvalues $-2l\hbar ,(-2l+2)\hbar ,\ldots
,2l\hbar $ and are disposed along the principal diagonal starting from the
lowest left corner. One should note that diagonal elements of the
perturbation $V$ vanish 
due to (\ref{hint}).
The global structure of the
matrix $H_{m,n}$ is shown in Fig.\ref{figband}. The next (to the principal
one) diagonals $H_{n,n\pm 1}$ correspond to transitions inside each $H_0$%
-shell while the ''arcs'' connecting the two corners represent transitions
between neighboring shells having $\Delta H_0=\pm 2\hbar $.
Such a global structure of the Hamiltonian matrix is not a
peculiarity of this model but it corresponds to the so-called ``shell
model'' representation widely used in atomic and nuclear physics \cite
{FGGK94,FBZ96}. It was shown in Ref. \cite{FI97,FI97a,BGIC97}
that,
generic properties of eigenfunctions
in this basis, can be directly related to
single-particle operators, in particular, with the distribution of
occupation numbers for single-particle states.

The Hamiltonian matrix has a clear band structure, with the bandwidth $b$
ranging from $1$ at the corners up to $b= 2l+1$ in the middle. However, this
structure differs strongly from that of standard Wigner Band Random Matrices
(see for example \cite{CCGI96,FCIC96} and references therein). Moreover,
non-zero off-diagonal matrix elements are positive and the mean and
variance of the distribution of these matrix elements depend on the
classical parameter $L^2 = \hbar^2 l (l+1)$ only. To be more precise, if one
assumes a continuous distribution of the matrix elements, it can be shown
(see Appendix 1) that $\sigma^2 = \langle v^2 \rangle - \langle v \rangle^2
\simeq (L/4)^4$.

There are of course semiclassical corrections to this estimate, but the
variation of $\hbar$ in one order of magnitude (available in our numerical
study) change the ratio $\sigma^2 / (L/4)^4 $ by less than 1 \%. In the same
way the mean value can be semiclassically estimated as $\langle v \rangle
\simeq (\pi L /8)^2 $. The agreement between these simple semiclassical
formulas and our numerical data is shown in Appendix 1.

\begin{figure}
\centerline{\epsfxsize=0.7\colwi\epsffile{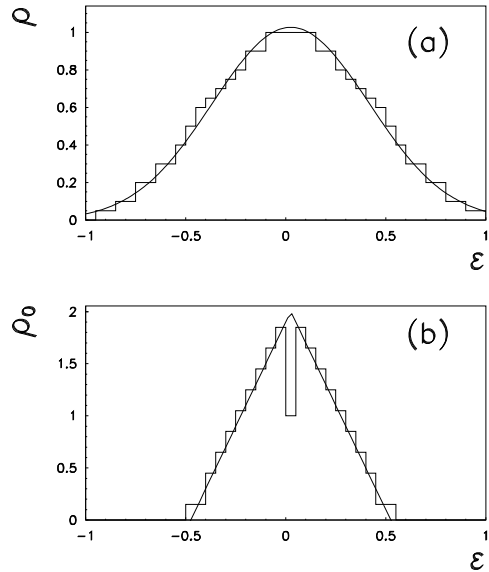}}
\caption{Density of states for the total Hamiltonian $H_0+V$, (a),
and for the unperturbed one $H_0$,  (b)
as a function of the rescaled energy $\epsilon = E/E_{max}$
for $l=39$ and $L=3.5$.
For comparison, the fitting
Gaussian (a) and the triangular curve (b) 
.}
\label{densi}
\end{figure}

The model is highly non-perturbative since 
the perturbation spreads
the levels
of the inner $H_0$-shell all over the allowed
energy range, see Appendix 2. As a result, the perturbed spectrum is broader
than the unperturbed one; this is an effect of the non-zero mean value of
off-diagonal matrix elements. While the unperturbed spectrum has
degenerate energy levels with spacing $\hbar$ (therefore, with density
of levels $\rho_0 \sim 1/\hbar$ and spectral radius $R^0_\sigma =
2\hbar l \sim 2 L $), the perturbed one has $(l+1)^2 $ non-degenerate
states within a spectrum of radius $R_\sigma \sim L^2 + 1$,
 which gives
the density $\rho \sim 1/\hbar^2 \sim \rho_0/\hbar $.

Numerical data show that the 
density of states changes from a triangular shape
for the unperturbed Hamiltonian (one more state is added/subtracted at each
neighbor level for $H_0$ negative/positive) to the Gaussian form
which is known to be generic for realistic finite systems like atoms and
nuclei, see for example \cite{FGGK94,FBZ96}. In Fig.\ref{densi} the
perturbed and the unperturbed density are shown for a typical case, together
with the corresponding fitting curves.

\section{Quantum-Classical Correspondence for EF and LDOS}

\label{sec:level2}

The subject of this section  is the analysis of the LDOS (also known as
''strength function'' or ''Green spectra'') and of the structure of
eigenfunctions, together with their classical analogs. In the quantum
description, all information is contained in the matrix constructed from the
eigenfunctions $\psi _n(E_m)$ of the total Hamiltonian $H$ represented in
the ordered unperturbed two-particle basis $|{\bf n}\rangle $. Here $\psi
_n(E_m)$ is the $n$-th component of the eigenfunction having $E_m$ as
eigenvalue. This matrix is assumed to be reordered in eigenenergies $E_m$.

In the classical limit the unperturbed energy $E_0$ is not constant when the
(chaotic) trajectory of the total Hamiltonian $H$ fills the $H = E = const $
surface. Indeed it fills a range of values which are distributed according
to  the ergodic measure on the constant energy surface, yielding a
distribution function $W(E_0 \vert E)$ \cite{CCGI96}. This distribution can
be easily numerically calculated taking a sample of chaotic trajectories $%
u(t) = ( L_x(t), L_y(t), L_z(t), M_x(t), M_y(t), M_z(t) )$ having the same
fixed values of $E$ and $L^2 = M^2$. Following these trajectories, one can
calculate $H_0 (u(t)) = L_z(t) + M_z(t)$ taken at equal instants of time and
find the distribution of $H_0$ over the energy band \cite{nota1}
defined by sojourn times.

The quantum analog of this distribution is provided by the relation

\begin{equation}
\label{LDOL}W_n (E) = \langle \vert \psi_n (E_m) \vert^2  \rangle_m  
\end{equation}

where the average $\langle\ldots\rangle$ is taken over those eigenfunctions
which have an eigenvalue $E_m$ in a fixed small energy interval around a
given energy $E$. Such an average has been done in order to smoothen the 
fluctuations 
which affect individual eigenstates; we would like to note that for
our dynamical model, unlike random matrix ensembles, there is no possibility
of ensemble averaging. The distribution $W_n (E)$ gives the average
shape of eigenstates represented in the unperturbed two-particles basis $%
\vert {\bf n} \rangle$.

In order to obtain the quantum distribution $W(E_0 \vert E)$ one needs to
switch to the unperturbed energy
representation, $n \to E_n^0$. Technically this can be realized by
introducing small energy bins $\Delta E$ and counting the correspondent
probability within them,

\begin{equation}
\label{LDOL1}W (E_0 \vert E) = \sum_n W_n (E) \delta (E_0 -E_n^0)  
\end{equation}

Similarly, we can define the distribution $w(E \vert E_0)$. In the quantum
case this distribution is the LDOS defined by

\begin{equation}
\label{LDO}w(E|E_0)=\sum_m\langle |\psi _n(E_m)|^2\rangle_n \delta (E-E_m)
\end{equation}
where the average is now taken over a number of values of $n$, such that the
eigenvalues $E_n^0$ belong to a small interval around the given unperturbed
energy $E_0$. The presence of degeneracy in this case provides 
an obvious  way of
taking  average. The corresponding classical function can be found by
noticing that the trajectory does not fill the whole surface $H_0=E_0$ but
is restricted to an invariant manifold specified by the value of $m$. 
Giving equal weight to all $m$ values corresponding to a given value of $H_0$%
, exactly matches the quantum averaging used in (\ref{LDO}). Then 
the classical distribution can be evaluated  analytically, since the
classical unperturbed Hamiltonian $H_0$ is integrable. Indeed, the
unperturbed solution $u_0(t)$ for $L^2=M^2$ and $H_0=0$ is given explicitly
by : 
\begin{equation}
\label{unpeq}
\begin{array}{ll}
L_x^0(t)= & \sqrt{L^2-m^2}\cos 2(t-\phi ) \\ L_y^0(t)= & \sqrt{L^2-m^2}\sin
2(t-\phi ) \\ L_z^0(t)= & m \\ 
M_x^0(t)= & \sqrt{L^2-m^2}\cos 2(t-\xi ) \\ M_y^0(t)= & \sqrt{L^2-m^2}\sin
2(t-\xi ) \\ M_z^0(t)= & -m
\end{array}
\end{equation}
where $0<\phi <2\pi $, $0<\xi <2\pi $ and $|m|\leq L$ depend on the initial
conditions. Therefore, 
$$
H(u_0(t))=(L^2-m^2)\cos 2(t-\phi )\cos 2(t-\xi ) 
$$
This means that the classical distribution of $H$ is given by $P_L(y)$ where 
$y=L^2(1-x_1^2)\cos \pi x_2\cos \pi x_3$ is a function of 
the random variables $%
-1<x_i<1$, $i=1,2,3$.

\vspace {-2cm}
\begin{figure}
\hspace{.8cm}
\epsfxsize 7cm
\epsfbox{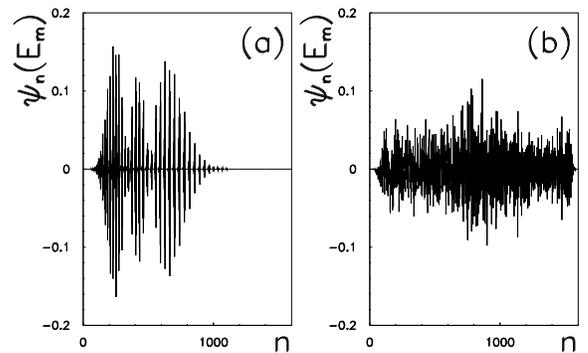}
\vspace{-2cm}
\narrowtext
\caption{Typical eigenfunctions  for
the case $L=3.5$, $l=39$. a) The second
excited state  $\psi_n(E_2)$ (corresponding to a
classically integrable region). b) Eigenfunction for
the energy $E_m$ close to zero
(middle of the spectrum, corresponding to the chaotic
region).
}
\label{exe}
\end{figure}

\section{Structure of eigenfunctions}

The quantum model has been already studied in Ref.\cite{FePeQ}, but
 previous studies have not addressed  the structure of eigenfunctions
in the two-body particle basis. This representation is quite natural and
corresponds to a well known procedure in the physics of interacting
particles.

In our dynamical model the structure of eigenfunctions strongly depends on
their energy because in the classical limit for low and high energies ($%
\vert \epsilon \vert = \vert E \vert /E_{max} \sim 1$) the motion is regular
while in the center of the energy band ($\vert \epsilon \vert \sim 0$) is
chaotic. One can, therefore, expect that in the classical limit ($\hbar \sim
L/l \ll 1$) the eigenstates corresponding to regular or chaotic regions are
very different. This is, indeed, clearly seen in Fig.\ref{exe} where two
eigenstates are plotted in the unperturbed two-particle basis for the second
(from the bottom of the spectrum) eigenstate ($\epsilon \sim -1$) and for
the eigenstate chosen in the center of the energy band ($\epsilon \sim 0$).
Comparing these two (typical) eigenstates, one can see that there are strong
correlations between components of the ``regular'' one (Fig.\ref{exe} a)
while the ``chaotic'' eigenstate can be treated as random along the whole
basis $\vert {\bf n}\rangle$. We also would like to note that the regular
eigenstate has many ``principal components'' and that it looks more or less
extended. This again indicates that the perturbation is quite strong and
effectively couples many unperturbed states.

A much more accurate analysis of eigenfunctions is obtained by studying
their localization lengths. Since the basis is finite and eigenstates can be
extended along it, here we use different measures of localization lengths,
based on their entropy ${\cal H}$ and participation ratio ${\cal P}$ see,
e.g. \cite{I90,FM94},

\begin{equation}
\begin{array}{rl}
l_H (E) = & 2.08 \ \exp \{ -
{\cal H} \} \\ l_{ipr} (E) = & 3 / {\cal P} \label{eql} 
\end{array}
\end{equation}

where 
$$
{\cal H } = \sum_{n=1}^N  \vert \psi_n (E) \vert^2 ln \vert \psi_n (E)
\vert^2 
$$
and 
$$
{\cal P} = \sum_{n=1}^N \vert \psi_n (E) \vert^4 
$$

The normalizing coefficients $2.08$ and $3$ were chosen 
 in order that $l_H =
l_{ipr} = N$ in the limit case when all components $\psi_n (E_m)$ are
independent Gaussian random  variables. Here $N = (l+1)^2$ is the size of
the two-particle basis.

\begin{figure}
\hspace{.8cm}
\epsfxsize 7cm
\epsfbox{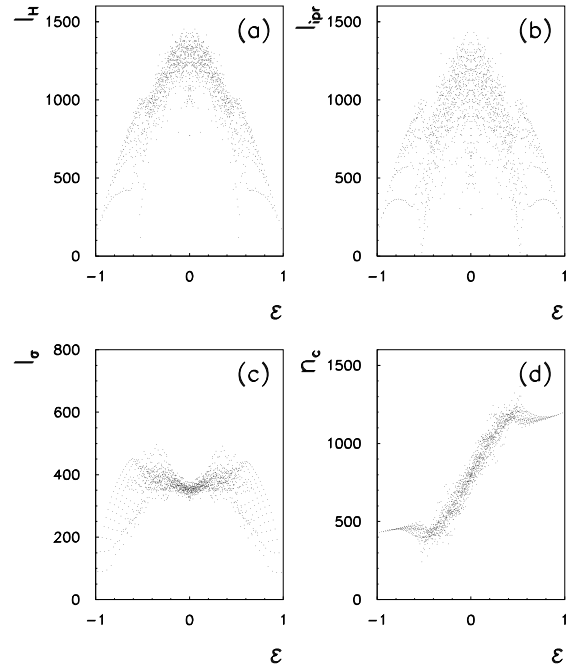}
\vspace{-0.5cm}
\narrowtext
\caption{Measures of localization lengths for eigenfunctions
versus the rescaled energy for
the case $L=3.5$, $l=39$. a) entropy localization length $l_H$,
 b) localization length $l_{ipr}$ defined as the inverse participation ratio,
 c) square root of the variance, $l_ \sigma$,
d) centroid $n_c$}
 \label{loc}
\end{figure}

Further information can be extracted from the centroids $n_c$ of eigenstates
and from their ``widths'':

\begin{equation}
\begin{array}{rl}
n_c (E) = & \sum_n n \vert \psi_n (E) \vert^2 \\ 
l_\sigma (E) = & \left( \sum_n \vert \psi_n (E) \vert^2 (n- n_c (E) )^2
\right)^{1/2} \label{eqlp} 
\end{array}
\end{equation}
In Fig.\ref{loc} we present numerical results for the above quantities as 
functions of the rescaled energy $\epsilon$.

First, we note that the entropy and inverse participation ratio localization
lengths, $l_H$ and $l_{ipr}$, are approximately equal and show the same
behavior, namely, the delocalization along the whole basis in the middle of
the spectrum, and the localization at the spectrum edges (see Fig.\ref{loc}
(a-b)). We also note that due to the underlying symmetry of the model the
above quantities $l_{ipr}$, $l_{{\cal H}}$ 
are symmetric around $\epsilon =0$.

Even in the center of the spectrum, where
eigenstates are in average maximally extended, there are big fluctuations in
the value of localization lengths. This indicates that in the classically
chaotic region there are some eigenstates which can not be treated as
completely random and delocalized over the energy shell. A careful study shows
that such eigenstates are characterized by an extended background with some
pronounced peaks (the so-called "sparse eigenstates"). 
Such eigenstates may result in the absence of equilibrium and in the lack of
standard statistical description, see details in \cite{FI97a}.

One can also see a clear regular structure in the dependence of $l_H$ and $%
l_{ipr}$ on the energy at the edges of the spectrum, which reflects the
regular character of eigenstates.

The other two quantities, $n_{c}, l_{\sigma}$ give an information about the
``position'' and ``width'' of eigenfunctions in the two-particle basis, see
Fig.\ref{loc}(c,d). In contrast to $l_H$ and $l_{ipr}$ the ``width'' $%
l_\sigma$ reveals a quite unexpected minimum at the center of the spectrum.
Additional numerical analysis shows that this is a result of different
"sparsity" of chaotic states depending on the energy. Namely, chaotic
eigenstates are more compact at the center of the energy band than far from
it. In fact, the ratio $l_\sigma/l_{H,ipr}$ can be used to extract an
information about the sparsity of eigenstates (see \cite{PR93}). Indeed, two
eigenstates with the same value of $l_{H,ipr}$ can have very different
values of $l_\sigma$ depending on whether principal components (those with
relatively large values of $\psi_n (E_m)$) are clustering around some center
(small $l_\sigma$) or randomly scattered over the whole unperturbed basis
(large $l_\sigma$).

Additional information can be obtained from the dependence of the centroids
of eigenstates on the energy, see Fig.\ref{loc} (d). Apart from fluctuations
and excluding the regular part, this dependence is linear, which means that,
in average,  centers of eigenstates are located at the center of the
energy shell covered by the classical distribution $w(E|E_0)$. This
generic feature of chaotic eigenstates has been studied in greater details
in WBRM models \cite{CCGI96}. In particular, it was shown that those
eigenstates, which are completely extended in the whole energy shell, are
characterized by maximal statistical properties of quantum chaos. For
example, in that case the statistics of energy spectrum follows the
predictions of Random Matrix Theory, such as the Wigner-Dyson form of the
distribution of spacings between neighboring energy levels. On the other
hand, 
localization of eigenstates within the energy shell  leads to
the so-called intermediate statistics \cite{I90} (which is
intermediate between the Wigner-Dyson and the Poisson statistics).
\vspace {-2cm}
\begin{figure}
\hspace{.8cm}
\epsfxsize 7cm
\epsfbox{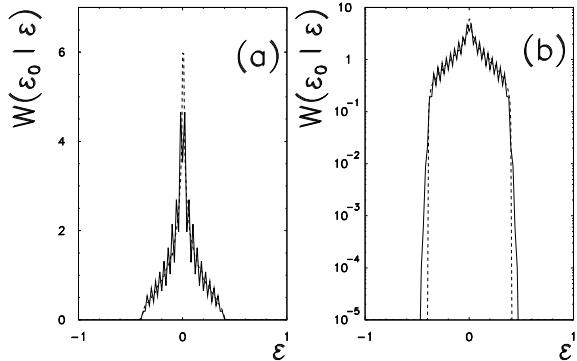}
\vspace{-2cm}
\narrowtext
\caption{a) Shape of eigenfunctions in the energy representation
(full line) and
classical distribution $W(\epsilon_0 \vert \epsilon)$
 (dashed)
for the case
$L=3.5$, $l=39$, obtained  by averaging over the central
$l + 1 $ eigenfunctions with $\epsilon_0 = 0$;
b) the same plot in semi-log scale.}
\label{ef}
\end{figure}

Such a localization is reflected in the fluctuations of $n_c $ around
the center of the shell (linear dependence on $\epsilon$). Indeed, if 
eigenstates are localized, their centers $n_c$ are typically scattered
within the energy shell leading to strong fluctuations of $n_c$;
 instead, this
cannot happen if they fill the whole energy shell. It is important to stress
that localization  in the energy shell is different from that  in the
unperturbed basis, as found from Eqs.(\ref{eql}).

The result presented in Fig.\ref{loc}(d) shows that localization in the
energy shell, if any, is quite weak. A more  direct 
analysis of the degree of
localization in the energy shell 
is provided 
 by the direct comparison of the
average shape of eigenstates in energy representation to its classical
analog. The results are presented in Fig.\ref{ef} where quantum and
classical  $W(E_0|E)$ (see Eq.\ref{LDOL1}) are plotted versus
the rescaled energy $\epsilon =E/E_{max}$.

One can see that the only important difference is a sort of a weak quantum
tunneling in the classically forbidden region (the tails of the classical
distribution are sharper than the quantum ones). Anyway, the good
correspondence between quantum and classical distributions shows that for
the chosen parameters the model is in a deep semiclassical region and,
globally, the eigenstates should be treated as ergodic ones (in the energy
shell, not in the whole unperturbed basis !). This means that the observed
scattering of the centroids of eigenstates (see Fig.\ref{loc}(d)) is, in
fact, quite weak and does not lead to noticeable localization in the energy
space.

Interestingly,
the size of these
ergodic eigenfunctions is smaller than the total energy band. The
distribution $W_0$ is infact 
restricted between the minimum and the maximum value
that the function $H_0 = L_z + M_z$ can assume 
 under the constraints $E=L_z +
M_z +L_x M_x$, $L_x^2 + L_y^2 +L_z^2 = M_x^2 + M_y^2 +M_z^2 = L^2$ . It can
be easily proved, using the Lagrange multipliers methods that for $E=0$, $%
\vert E_0 \vert < 2( \sqrt{L^2+1} -1 )$, which in Fig.\ref{ef} corresponds
to $\vert Supp W(\epsilon_0 \vert \epsilon) \vert < \tilde{\epsilon} \simeq
0.398$

\vspace {-2cm}
\begin{figure}
\hspace{.8cm}
\epsfxsize 7cm
\epsfbox{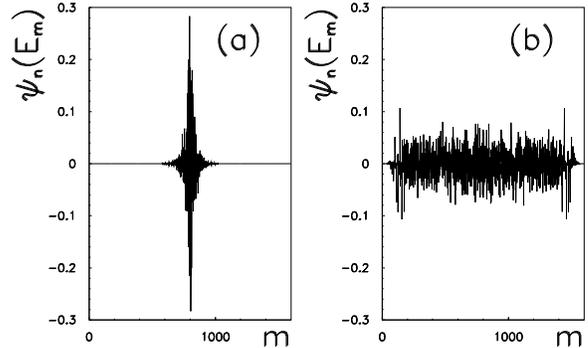}
\vspace{-2cm}
\narrowtext
\caption{
a) The BS for an $n$ value  corresponding to a shell edge
indicated as a dashed line in Fig.(8).
b) The BS in the middle of $H_0 = 0$ shell.}
\label{exe1}
\end{figure}

\section{LDOS structure}

Of special interest is the structure of LDOS which is widely discussed in
many applications in atomic, nuclear and solid states physics. The
importance of this quantity relates to its physical meaning: it shows how an
unperturbed state $\vert {\bf n}\rangle$ "decays" into other states due to
interaction. In particular, the inverse width of the LDOS is associated
with the mean "lifetime" of a chosen basis state.

As was indicated above, the LDOS structure can be extracted from the matrix $%
\psi_n(E_m)$ by fixing an unperturbed state $\vert {\bf n}\rangle $ and
searching the dependence on $m$. Therefore, we can adopt the same procedure
as we did above when analyzing the structure of eigenstates. In comparison
with Fig.\ref{exe} we show two such ``matrix lines'' corresponding to 
basis states (BS), with close $n$ values, taken from the center of matrix (see
Fig.\ref{exe1}). In fact, the BS-lines in this matrix correspond to the
expansion of unperturbed (basis) states in the exact eigenstates. While the
structure represented in Fig.\ref{exe1}(b) is typical, the form of the BS in
Fig.\ref{exe1}(a) is only observed around some specific value of $n$.

\begin{figure}
\hspace{.8cm}
\epsfxsize 7cm
\epsfbox{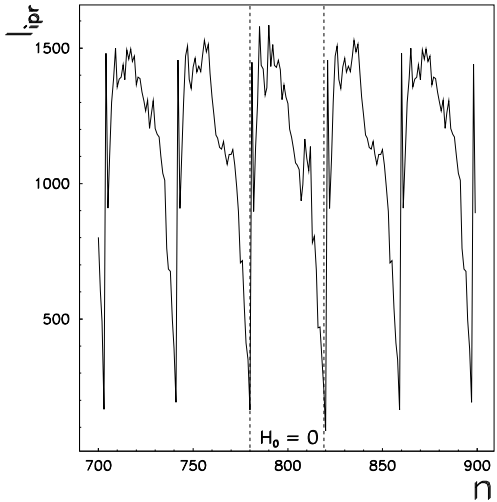}
\vspace{-1cm}
\narrowtext
\caption{Dependence of the inverse participation ratio
of the $n-$th BS of the $H$ matrix as a function of $n$.
Here is $L=3.5$, $l=39$.
The $H_0 = 0$ shell
is inside the two vertical dashed lines.}
\label{period}
\end{figure}

To better understand the meaning of these peculiar $n-$values we have
computed the localization length (compare with Eq.(\ref{eql}))

\begin{equation}
\begin{array}{rl}
l_{ipr} (n) = & 3 / \sum_m \vert \psi_n (E_m) \vert^4 \label{eql1} 
\end{array}
\end{equation}
in some range of $n$. The data presented in Fig.\ref{period} reveal a global
periodic structure of BS, from which one understands that the 
peculiarity of BS reflected in Fig.\ref{exe1} results from the degeneracy
(inside each shell) of the unperturbed spectrum. It is then convenient to 
consider 
in the following analysis, as a reference, the central shell $H_0=0$ only
(set of $l+1$ BS ).

\vspace {-1cm}

Now we discuss the structure of the BS in energy space which is, in fact,
the LDOS (Eq.\ref{LDO}). According to results\cite{FCIC96} obtained for
WBRM, the shape of LDOS typically changes from the Breit-Wigner (BW) law to
the semicircle when an effective perturbation is increased. In particular,
the BW is expected when 
\begin{equation}
\label{esti}1/\sqrt{2\pi} \ll \rho_0 V \ll \sqrt{b/2\pi}  
\end{equation}
where $\rho_0$ is the density of the unperturbed spectrum and $b$ is the
effective band width of the Hamiltonian matrix. The first inequality is
related to the non-perturbative character of the coupling (which is always
verified in this model) while the last, rewritten as 
$$
2\pi \rho_0 V^2 = \Gamma_F \ll b/\rho_0, 
$$
simply means that the spreading width $\Gamma_F$ of such distribution has to
be much smaller than the energy bandwidth $b/\rho_0$. If we formally apply
the above conditions to our case, we get (in units of $\epsilon = E/E_{max}$%
) the following relations: $b/\rho_0 \simeq 1/L $ and $2\pi\rho_0 V^2 \simeq
(\pi/2) (L/4)^3 l $. With our
 data the second
condition in Eq.(\ref{esti}) is strongly violated. 
In fact the random matrix agrgument leading to the above results
rests on the assumption that the band in the Hamiltonian matrix be ``full'';
in our case, instead, we have a large sparsity (many vanishing 
matrix elements inside the band).

\begin{figure}
\vspace {-2cm}
\hspace{.8cm}
\epsfxsize 7cm
\epsfbox{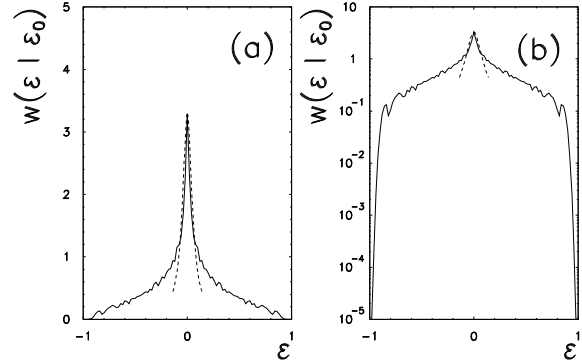}
\vspace{-2cm}
\narrowtext
\caption{a) Quantum LDOS distribution $w (\epsilon \vert\epsilon_0)
= w(E/E_{max}\vert  E_0/E_{max})$
(full line) and best fitted Breit-Wigner
distribution in the range specified above (dashed line) for the case
$L=3.5$, $l=39$, obtained averaging over $l+1$ values of BS for the $H_0=0$
shell; b) the same plot in semi-log scale.}
\label{BW}
\end{figure}

In Fig.\ref{BW} we show the structure of the LDOS for the BS corresponding
to the center of the unperturbed spectrum, in comparison with the
Breit-Wigner fit which is performed within the interval $(-b/2\rho_0,
b/2\rho_0)$.

\vspace {-2cm}
\begin{figure}
\hspace{.8cm}
\epsfxsize 7cm
\epsfbox{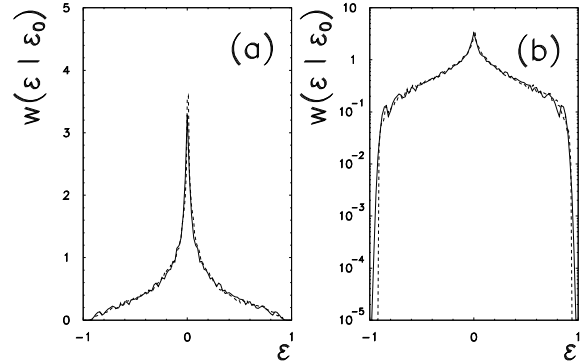}
\vspace{-2cm}
\narrowtext
\caption{a) The LDOS distribution $w (\epsilon \vert \epsilon_0)$
: quantum (full line) and
classical (dashed line)
for the case
$L=3.5$, $l=39$, obtained  by averaging over $l + 1$ values of BS for $H_0=0$
shell;  b) the same plot in semi-log scale.}
\label{LDOS}
\end{figure}

One should stress that outside of the energy interval corresponding to the
band size, the tails of LDOS are known to be highly non-generic \cite
{FGGK94,FCIC96,S96,FBZ96} depending on specific properties of the model. As
one can see, inside this energy interval the shape of LDOS can be roughly
associated with the BW form. On the other hand, outside, the tails decay
very slow, compared to those given by the BW. Such a form of the tails is
also different from the case of the WBRM \cite{FGGK94} where outside the
band energy range the tails decay extremely fast (even faster than
exponential). In general, the above results seem to indicate that the
effective perturbation corresponds just to the condition when the BW
approximation starts to fail.

Let us now compare quantum and classical LDOS. In Fig.\ref{LDOS} we give an
example of such distributions in the whole energy shell.
They coincide with a high accuracy, apart from the regions
very close to the energy shell edges (where quantum tunneling
is significant).

This again means that the system is in a deep semi-classical regime. 
We remark that, 
in this model, 
going to the quantum regime in the chaotic energy
region calls for  very small matrices,
 for which fluctuations are extremely
strong.

\vspace {-2cm}
\begin{figure}
\hspace{.8cm}
\epsfxsize 7cm
\epsfbox{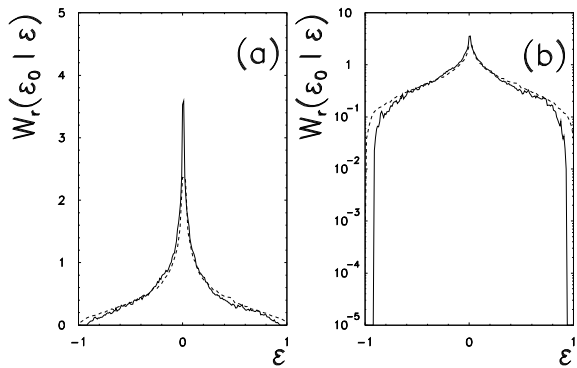}
\vspace{-2cm}
\narrowtext
\caption{a) Classical LDOS distribution $w (\epsilon, \epsilon_0)$
 (full line )
and the rescaled classical distribution $W_r (\epsilon_0\vert\epsilon)$
for the case
$L=3.5$, $l=39$.
b) the same plot in semilog scale.}
\label{resca}
\end{figure}

An important question is the relevance of the shape of eigenstates to that
of the LDOS. As was shown in the model of WBRM \cite{CFI97} in some range of
parameters (for not very strong perturbation) the two shapes are very close
to each other, which is a manifestation of the ergodic structure of
eigenstates in the energy shell. On the other hand, with an increase of
perturbation the LDOS was found to tend to the semicircle, for which strong
localization turns out to be possible \cite{CCGI96}. This localization
manifests itself in different average shapes of the EF and LDOS. Namely, the
width of the EF in the energy representation is less that the width of the
LDOS; the latter defines, in fact, the width of the whole energy shell.

Direct comparison of Fig.(\ref{ef}) and Fig.(\ref{LDOS}) in our dynamical
model shows a remarkable different energy range for the LDOS and
EF distribution. As was discussed above, the energy width of
eigenfunctions in the semiclassical region (in the energy representation),
is much smaller than the width of the spectrum because it is subject to an
additional constraint. We can take into account this restriction and rescale
the distribution $W(\epsilon _0|\epsilon )$ in order to have the same energy
range as for $w(\epsilon |\epsilon _0)$: $W_r(\epsilon _0|\epsilon )=\tilde
\epsilon W(\epsilon _0/\tilde \epsilon |\epsilon /\tilde \epsilon )$ where $%
\tilde \epsilon =2(\sqrt{L^2+1}-1)$. The rescaled distribution $W_r$ is
presented in Fig.\ref{resca} together with the distribution $w(\epsilon
|\epsilon _0)$. After such a rescaling both distributions coincide quite
well which again indicates the absence of the localization.

\vspace {-2cm}
\begin{figure}
\hspace{.8cm}
\epsfxsize 7cm
\epsfbox{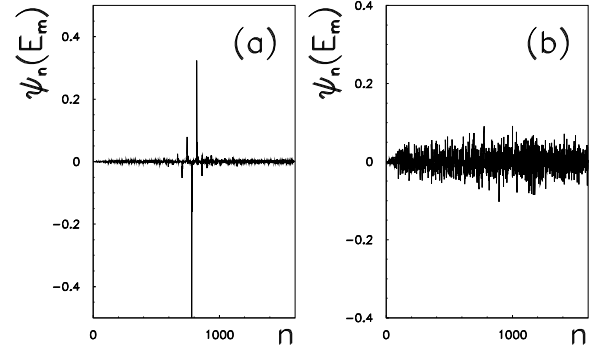}
\vspace{-2cm}
\narrowtext
\caption{
Typical eigenstates
for the case $L=3.5$ and $l=39$ and
Gaussian random non-zero off-diagonal elements.
(a) and (b) have close eigenvalues in the middle of the spectrum,
respectively $\epsilon = E/E_{max} = 1.48 \cdot 10^{-3}$ and
$\epsilon = E/E_{max} = 2.295 \cdot 10^{-2}$  but very different
inverse participation ratios ( $l_{ipr}/N= 0.0027$ for the first and
$l_{ipr}/N= 0.87$ for the last, where $N$ is the matrix size.)
}
\label{exer}
\end{figure}
\section{Random two-body interaction}

\label{sec:level4}

In this Section we modify our  model of two interacting particles 
by assuming a
completely random interaction which preserves some global properties of the
original dynamical model (\ref{ham1}). Namely, the unperturbed part $H_0$ is
taken exactly the same as in the dynamical model. However,
we replace non zero matrix elements of the dynamical model with 
 random and independent variables.
Moreover, we choose
a Gaussian distribution of these 
random matrix elements  with the same mean and
variance
as for the dynamical elements.
In such a way we can reveal the influence of dynamical
correlations which are due to the specific form of the interaction $V$.
Below we follow the same procedure described in previous sections when
studying eigenstates and  LDOS.

Numerical data for the ``randomized" model show that global spectral
properties are the same as in the dynamical model. Namely, the perturbed
spectrum is enlarged with respect to the unperturbed one and the density of
states keeps the same Gaussian shape with the same mean and variance.

On the other hand, the analysis of eigenfunctions reveals clear differences.
Typical shapes of eigenstates and BS are shown in Fig.\ref{exer}. Comparing
with the corresponding Fig.\ref{exe}(b), one notes that extended states
look chaotic, similar to those found in the dynamical model.
However,  differently from the dynamical model, a 
few strongly localized states
now appear even in the center of the energy band.
A typical example of
such an eigenstate is given in Fig.\ref{exer}(a).

\begin{figure}
\hspace{.8cm}
\epsfxsize 7cm
\epsfbox{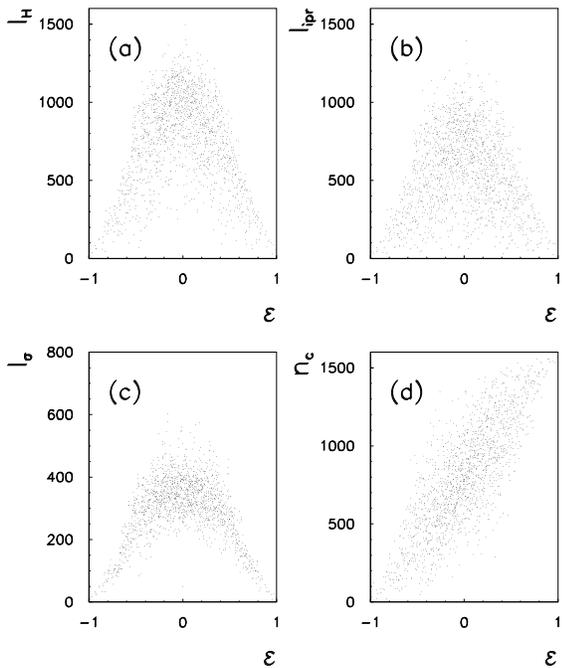}
\vspace{-0.5cm}
\narrowtext
\caption{Measures of localization lengths for eigenfunctions
for the Gaussian random case $L=3.5$, $l=39$; compare with Fig.(5).
}
\label{locr}
\end{figure}

To analyze the global characteristics of all eigenstates we have calculated
different localization lengths $l_H$ and $l_{ipr}$ as well as the width $%
\l_\sigma$ and the centroids $n_c$ according to Eq.(\ref{eql},\ref{eqlp}).
The data reported in Fig.\ref{locr} should be compared with those in Fig.\ref
{loc} for the dynamical model. As expected, for the random model there
are no correlations in the energy dependence for large/small energy ($\vert
\epsilon\vert \simeq 1$), compared with Fig.\ref{loc}. However, close to the
edges of the energy spectrum, the eigenstates can not be treated as chaotic
since the number of ``principal components'' in such eigenstates is quite
small (this is revealed by small values of localizations lengths $l_H,
l_{ipr}, l_{\sigma}$). This is a result of the perturbative localization
which typically occurs for states close to the ground state. In what
follows, we exclude such states from our consideration.

For chaotic eigenstates, the various measures of localization lengths give
average values typically less than in the dynamical case. This holds
especially in the middle of the spectrum $E \simeq 0$ where we can now find
a relatively large number of sharply localized eigenstates (with $l_{ipr}/N
\ll 0.1$ where $N$ is the matrix size), see Fig.\ref{exer} (a). In the same
Fig.\ref{locr}(d) one can also observe a much more stronger scatter of the
centroids of eigenstates transverse to the diagonal (compare with Fig.\ref
{loc}(d) of the dynamical model). The same features have been found for
basis states. These data indicate that fluctuations in the structure of
eigenstates are much stronger than in the dynamical model.
\vspace {-2cm}
\begin{figure}
\hspace{.8cm}
\epsfxsize 7cm
\epsfbox{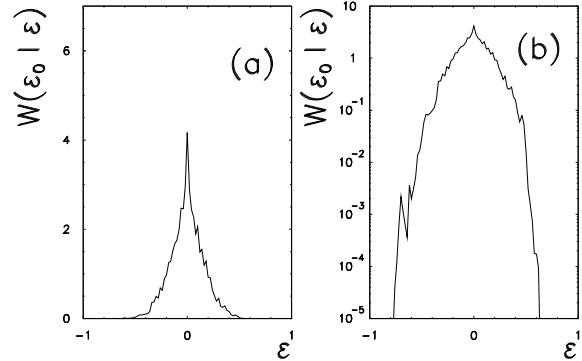}
\vspace{-2cm}
\narrowtext
\caption{
(a) The EF distribution for the Gaussian random case with $L=3.5$, $l=39$,
obtained by averaging over $l+1$ central eigenfunctions;
(b) the same as (a) in semilog scale.
}
\label{efr}
\end{figure}

\begin{figure}
\hspace{.8cm}
\epsfxsize 7cm
\epsfbox{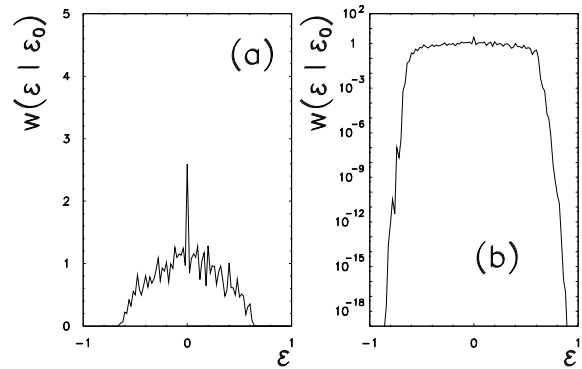}
\vspace{-2cm}
\narrowtext
\caption{ (a) The LDOS distribution for the Gaussian random
case with $L=3.5$, $l=39$,
obtained by averaging over $l+1$ central lines;
(b) the same as (a) in semilog scale. }
\label{ldosr}
\end{figure}

Despite these fluctuations, the global structure of the EF seems to remain
the same. This is marked once more by the distribution of eigenfunctions in
the energy space (the analog of Fig.\ref{ef} is now shown in Fig.\ref{efr})
which has the same shape as in the dynamical model. However, the LDOS
distribution for the random model shows striking difference, see Fig.\ref
{ldosr}. Indeed it can be described, apart from the central peak, by the
semicircle law \cite{W55,FCIC96,CCGI96}.
\vspace {-2cm}

This surprising result 
is quite significant 
in the light of application of
Random Matrix models. Indeed, in \cite{CCGI93,CCGI96} it was found that
localization in energy shell for the WBRM may occur only when the LDOS is
characterized by the semicircle law. Therefore, the important question is
whether the semicircle law is a quantum feature or it can also occur in
classical dynamical conservative systems. What we have found here
is that the semicircle law has nothing to do with the semiclassical limit in
our model. It seems 
to be dictated by quantum randomness
rather than by the 
pseudo-randomness resulting from the classical chaos.

By comparing the shapes of the LDOS and EF for the random model, one can see
that they are clearly different, in contrast to the dynamical model. As 
mentioned 
above, in Ref. \cite{CCGI96} such a 
difference was directly connected to
the localization of eigenstates in the energy space. As a result of this
localization, the spectrum statistics differs from that predicted by the
RMT. In particular for Wigner Band Random Matrices, the level spacing
distribution was found \cite{CCGI96} to deviate from the Wigner-Dyson
dependence.

To check these predictions, we have calculated the level spacing
distribution for both dynamical and random models for the part of the
spectrum corresponding to chaotic eigenstates. As we expected, for the
dynamical model we have observed a very good correspondence to the
Wigner-Dyson dependence. Surprisingly, we have found that random model gives
the same result. This means that the level spacing distribution is quite
insensitive to the 
small
number of localized eigenstates. On the other hand, this result indicates
that,
in the case of realistic matrices, the degree of level repulsion
is not so clearly affected by the 
difference in the shapes of the LDOS and EF,
as it was in the case of WBRM.

\section{Summary}

\label{sec:level5}

In this paper we have studied a dynamical model with two interacting particles
(rotators). The classical version of this model manifests both regular
and chaotic motion, depending on the total energy: both at small and at large
energies the motion is regular, while, at intermediate energies, chaotic
properties are very strong. The quantum analog of this model can be assumed
 to
describe two interacting spins.
 Our choice was restricted to the subset of
symmetric states, which corresponds to particles with integer spins.

This model has already been under investigation, both
in the classical and in the quantum description (see, for 
example Ref.\cite{FePe}).
However, here we have used an approach which seems much more instructive:
in the quantum case, we have represented the Hamiltonian matrix
in the basis defined by the two-body eigenstates of the 
non-interacting system, 
 reordered 
according to increasing total 
energy. Such a representation corresponds to a well known procedure in
atomic and nuclear physics (``shell-basis representation''), and seems to be
very useful in view of recent developments \cite{FI97,FI97a}.

In this representation the Hamiltonian matrix turns out to be banded, with
many zero elements inside the band. If pseudo-randomness of 
non-zero off-diagonal elements is assumed (in the region of classical chaos), 
then one
can refer to
 some modern developments of Random Matrix Theory: in particular, 
to the
so-called Wigner Band Random Matrix ensemble, which is conjectured 
to be  well suited to the
description of conservative systems with complex behavior (see \cite{CCGI96}
and references therein). However, the assumption of pseudo-randomness 
of matrix elements is far from obvious: checking it was in fact one of the 
major motivations of our work. 

Random matrices 
in the WBRM  ensemble are characterized by a sharp
band inside which matrix elements are random, independent, and 
identically distributed,  
plus an
additional principal diagonal 
with increasing entries, corresponding to the
unperturbed spectrum of the two-body Hamiltonian.

Compared  to WBRM, our model has two peculiarities. In the first place
there is no free parameter of interaction between the particles,
 the only
parameter which determines the relative strength of the interaction being the
total energy of the system.
Second, 
our model has an 
highly degenerate  unperturbed
spectrum.
Still, the main 
features of the model are  expected to be quite generic, because
these peculiarities 
are quite typical in such  
physical applications as complex atoms and nuclei. 

In this paper
we have analyzed two main issues 
, motivated by recent results \cite{CCGI96,FI97,FI97a}.
First, 
we have studied 
the structure of 
the eigenfunctions and of the 
local density of states (LDOS)
and have compared them 
to 
 what is known 
for completely random models, and for  WBRM in 
particular. 
Second, 
we have looked for effects of dynamical localization; though we have found 
no significant evidence for such effects, our analysis has 
brought into light 
a close connection (surmised in \cite{CCGI96})
between  LDOS, eigenfunctions, and certain classical 
distributions,  which can be easily found by solving 
the classical equations of motion. 
 
As expected, in the region of classical  regular motion,
eigenstates have a regular 
structure themselves; 
still, classical integrability 
does not result in strong
localization, because 
these eigenstates are typically 
quite extended over  the basis of
two-particle unperturbed states.
On the contrary,  in the region of classical chaos,
the structure of eigenstates 
looks very chaotic itself.  Nevertheless,
the
size of such chaotic eigenstates  can be smaller
than the size of the basis, though the eigenstates may be treated as random
(ergodic) ones on the scale of their localization.

The dependence of the structure of eigenstates 
on energy reflects their regular or chaotic 
nature, as it is chaotic itself in the latter case;
e.g., fluctuations of the number of principal components
are stronger where classical chaos is stronger, which  is the reason why a 
nonnegligeable fraction of eigenstates have a size significantly 
smaller than the basis size.

Generally speaking,  
the global properties of chaotic eigenstates 
are quite similar to those found for 
WBRM, with one remarkable exception.
In fact,  
for the WBRM ensemble, 
the expansion of exact eigenstates
over the unpertubed ones 
has a structure 
quite similar 
to the one 
observed on expanding 
unperturbed eigenstates on the 
basis of exact eigenstates.
Instead, this symmetry is broken in our model,
apparently due to the degeneracy of the unpertubed spectrum:
a feature which is missing in WBRM.

Expansion of unpertubed eigenstates 
on exact ones directly leads to LDOS. 
In standard random matrix models the latter is known 
to be of the
Breit-Wigner (BW) type, with the half-width given by the Fermi Golden Rule. 
Instead, in our dynamical model 
the LDOS is BW--like  
only around the central peak;  its tails 
have a much slower decay than predicted by the BW law.

Dynamical localization effects are an extremely important issue 
when investigating the quantum mechanics of chaotic systems.
Our approach to this problem was based on  
Ref.\cite{CCGI96}, where it was argued  
that, for the case of conservative systems, such 
effects are manifested by localization of eigenstates 
within the so-called energy shell, which is 
the range of energies ergodically explored by classical motion.
From this viewpoint, in order to detect localization (if any),
one has to find the form of the classical energy shell, and then 
to compare it with the form of chaotic
quantum eigenstates.

Following this approach,
we have defined and numerically computed classical distributions
which strikingly correspond to the LDOS and to the average shape
of eigenfunctions . 
Beside opening a new direction 
in the study of the global properties of quantum LDOS and EFs,
this fact may be important 
for the quantum statistical mechanics of isolated, chaotic systems
 of interacting particles, because
the knowledge of the average shape of eigenstates 
gives analytical access to the 
distribution of occupation numbers of single particle states\cite{FI97,FI97a}.
However, insofar as localization effects are concerned, 
the close agreement we have observed 
between  quantum and classical distributions
indicates that no such effect is present in our model,  
which is, in fact, in a deep quasi--classical region. Additional 
indications of absence of 
significant localization effects is provided by the analysis of 
the level-spacing distribution, which  closely follows, in the strongly 
chaotic case, the predictions of Random Matrix theory.  

Finally, we have studied a Random Matrix analog of our dynamical
model, 
which was constructed by leaving the unperturbed part 
of the Hamiltonian matrix unaltered and by replacing 
all non zero off--diagonal
elements by Gaussian  random variables with the same mean and
variance as in the dynamical model.
In this way we were able to check to what extent quantum chaotic 
 dynamics can be simulated by random interactions; in other 
words, we have checked 
the pseudo-randomness assumption.
We have found that  the random matrix model and the dynamical one 
are very similar in what concerns the global average properties of
eigenstates.
Nevertheless we have found that fluctuations of individual
eigenstates are significantly stronger in the random model:
in particular there are many more eigenstates which 
are significantly more localized in comparison to the average
size of chaotic eigenstates. In spite of this 
enhancement of the number of localized states, 
the level spacing distribution of the random model 
is still short of showing significant deviations from Random Matrix Theory.

The most striking difference between the dynamical and the 
random model has been detected in the form of the LDOS.
The LDOS of the random model drastically differs from that of the
dynamical model, as it is 
quite close to the semicircle law, with an additional
peak at the center. While
the origin of this peak is 
related to a 
specific feature of our model,
the
occurrence of the semicircle 
is somewhat surprising, because 
the general statistical properties 
of the random model are similar to those of 
the dynamical one.
For WBRM, the semicircle law appears when the perturbation
(that is, the variance of the off-diagonal elements) is strong;
moreover, localization in the energy shell was found to
appear only in the presence of the semicircle law. 
In contrast to the dynamical model, 
neither WBRM nor the random model have
a classical analog (although the latter is much closer to
a realistic systems than WBRM); therefore one can ask the question,
whether the semicircle law for the LDOS can appear
at all in quantum systems with a chaotic classical limit. 
Our analysis 
shows that  
great care has to be taken in extending  
predictions of Random Matrix Theory to systems of 
the latter class, at least if the systems themselves are in a quasiclassical 
regime. In that case the pseudo-randomness assumption 
obliterates dynamical correlations to which LDOS, and similar quantities,  
are quite sensitive.

\section{Acknowledgments}

(FMI) thanks
with pleasure the colleagues of the University of Milan at Como for the
hospitality during his visit when this work was done; he acknowledges the
support from the Grant of the Cariplo Foundation for Research and partial
support from the INTAS Grant No. 94-2058.

\section{Appendix 1}

\label{sec:level6}

In this appendix we show that assuming a continuous distribution of
off-diagonal non zero elements, average and variance can be estimated
semiclassically and good agreement with numerical data is found. Let us
start with Eq. (\ref{hint}) from which we have : 

\begin{eqnarray}
\langle v \rangle  & = &  { {\hbar}^2 \over {M^2} } \sum_{i=1}^M \sum_{j=1}^M
\left[ (i+l) (i-l+1) (j+l) (j-l+1) \right]^{1/2} \nonumber\\
& \simeq  &   { {\hbar^2} \over {16 l^2} }
\int_{-l}^l \ dx \ \sqrt{ (x^2 - l^2)}
\int_{-l}^l \ dy \ \sqrt{ (y^2 - l^2)}
\label{av}
\end{eqnarray}

where, as usual $L^2 = \hbar^2 l(l+1)$. Integrals can be easily evaluated
and one has : 
\begin{eqnarray}
\langle v \rangle \simeq \left({ { \hbar l \pi }\over {8} } \right)^2
\simeq  \left({ { \pi L }\over {8} } \right)^2
\label{av1}
\end{eqnarray}

In the same way : 

\begin{eqnarray}
\langle v^2 \rangle & = & \left( { {\hbar^2}\over {4 M} }\right)^2
\sum_{i=1}^M (i^2 - l^2) \sum_{j=1}^M (j^2 - l^2)  \nonumber\\
\simeq 
& & \left( {{\hbar}\over{2} } \right)^4
\left[ {{1}\over{2l}} \int_{-l}^{l} dx (l^2 - x^2) \right]^2 = \nonumber \\
& = &  \left( { {\hbar^2 l^2} \over{6} } \right)^2 \simeq
\left( { {L^2} \over{6} } \right)^2
\label{av2}
\end{eqnarray}

in such a way that 
\begin{equation}
\label{avs}\sigma^2 = \langle v^2 \rangle - \langle v \rangle^2 \simeq
\left( {\frac{ {L} }{{4} }} \right)^4  
\end{equation}

\begin{figure}
\vspace {-2cm}
\hspace{.8cm}
\epsfxsize 7cm
\epsfbox{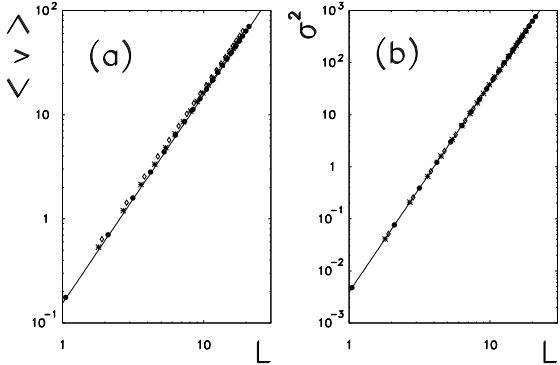}
\vspace{-2cm}
\narrowtext
\caption{ a) Off diagonal non-zero matrix elements average
as a function of classical $L$ and $l=9$ (circles),
$l=19$ (squares), $l=39$ (crosses); line is
the semiclassical expression
$(\pi L/8)^2$.
b) Off diagonal non-zero matrix elements variance
as a function of classical $L$ and $l=9$ (circles),
$l=19$ (squares), $l=39$ (crosses);
line is
the semiclassical expression
$(L/4)^2$.
}
\label{figapp}
\end{figure}

The agreement between Eqs.(\ref{av1}),(\ref{avs}) and numerical data is
shown in Fig.\ref{figapp}.

\section{Appendix 2}

\label{sec:level7}

It is instructive to estimate the splitting of the energy levels within one $%
H_0$-shell due to the perturbation, using degenerate perturbation theory.
Let us consider, for instance, a shell with $H_0 = 2\hbar j > 0$ which has
degeneracy $p = l+1-j$. Perturbed energy levels can be calculated by
diagonalizing the matrix:

\begin{equation}
\label{pert}h_{s,s^{\prime}} = \langle s, 2j-s \vert V \vert s^{\prime},
2j-s^{\prime}\rangle  
\end{equation}
This is a symmetric tridiagonal matrix with zero elements along the
principal diagonal, whose elements for any $s \ge 2$ are given by 
\begin{equation}
\label{pert1}h_{s,s+1} = {\frac{{\hbar^2}}{{4}}} \sqrt{[ (l-s+1)^2
-j^2][(l+s)^2 -j^2]}  
\end{equation}

The distance between two neighboring perturbed levels can be estimated as
the difference between two neighboring matrix elements (due to the symmetry
of the matrix),

\begin{equation}
\label{pert2}h_{s+1,s+2}-h_{s,s+1} \sim {\frac{ {\hbar^2}}{{2} }} l  
\end{equation}
where the approximation is taken for $j=s=0$. This means that the total
splitting is of the order

\begin{equation}
\label{stima}\Delta E \sim 2 p l\hbar^2/2 \sim \hbar^2 l^2 \sim L^2  
\end{equation}

\begin{figure}
\centerline{\epsfxsize=0.7\colwi\epsffile{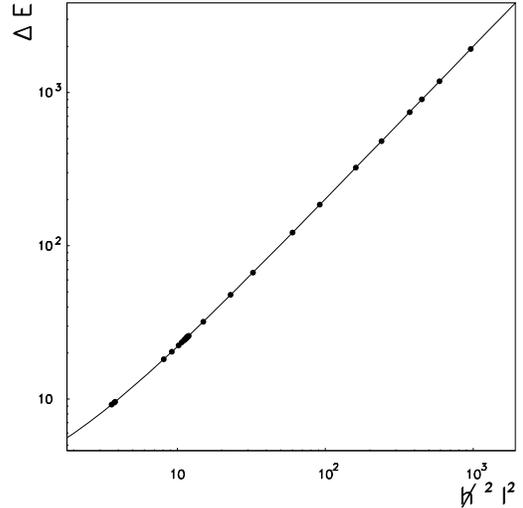}}
\caption{Energy spectrum splitting $\Delta E$
due to the perturbation $V$, as a function
of $\hbar^2 l^2$ for $0.01 < \hbar < 2$, $l<40$ and $L>1$ (points).
Full line is the classical energy shell  $\Delta E = 2 (L^2 + 1 ) $
.}
\label{deltae}
\end{figure}

Though this approximation is obtained for $j=0$, corresponding to the
biggest $H_0$-shell, similar behavior is expected for other shells. The
expression (\ref{stima}) has been checked numerically, see Fig.\ref{deltae},
where we plot $\Delta E=E_u-E_l$ as a function of $\hbar ^2l^2$ for
different $l$ and $\hbar $ (here $E_u$ and $E_l$ are the energy of the upper
and lower split level within one $H_0$-shell) For the comparison, in Fig.%
\ref{deltae}, the relation $\Delta E=2(1+\hbar ^2l^2)$ is also shown, which,
in the classical limit $\hbar \to 0,\quad l\to \infty $, and for $L>1$
yields the expression for the classical energy shell, $\Delta E=2\
E_{max}=2(1+L^2)$.

\begin{references}
\bibitem{nota1} 
Numerical results are obtained within the standard numerical
accuracy, that is they are independent of the integration time
and of the number of chaotic trajectories.
\end{references}
\end{document}